\documentclass
[prl,aps,twocolumn,superscriptaddress,showpacs,tightenlines,amsmath,amssymb]
{revtex4}

\usepackage{amssymb}
\usepackage{graphicx}
\usepackage{subfigure}

\begin{document}

\title{Analytical solution of a double-well Bose-Einstein Condensate}

\author{Carlos Sab{\'\i}n}
\address{School of Mathematical Sciences, University of Nottingham, Nottingham NG7 2RD
United Kingdom}
\author{Pablo Barberis-Blostein}
\address{Instituto de Investigaciones en Matematicas Applicadas y Sistemas, Universidad Nacional Aut\'onoma de M\'exico}
\author{Ivette Fuentes}
\address{School of Mathematical Sciences, University of Nottingham, Nottingham NG7 2RD
United Kingdom}
\date{\today}
\begin{abstract}
We introduce a microscopic computation which shows that the Hamiltonian of a Bose-Einstein Condensate can be analytically solved in the two-mode approximation, in particular, in the case of an asymmetric double-well condensate in the dilute regime. Our model is exactly diagonalisable when the overlap of the quasilocalized modes in each well is small enough with respect to the trap asymmetry. For larger overlaps or highly symmetric traps, our diagonalisable Hamiltonian acquires extra terms that we treat within perturbation theory.
 
\end{abstract}
\pacs{03.75.Gg, 42.50.Gy,03.75.Lm}
\maketitle

The Hamiltonian of a two-mode Bose-Einstein condensate (BEC) has been extensively studied in the literature \cite{cirac,internal,theoryspa,MIT,JILA2} due to its relative simplicity and application to double-well \cite{MIT} or spin-$1/2$ \cite{JILA2} condensates. 

In spite that the two-mode BEC corresponds to the most simple multi-mode situation, canonical models  employed, such as the Bose-Hubbard Hamiltonian \cite{cirac}, cannot be solved analytically. Therefore, common techniques to study this system involve numerical methods, approximations (such as the use of Bethe ansatz) and semiclassical analysis (mean-field theory) where the modes are treated classically.  However some of the authors of this paper have introduced a model which Hamiltonian can be diagonalized analytically \cite{us,us2}. The model is a generalization of the Bose-Hubbard Hamiltonian which includes mode-exchange and coherent tunneling interactions previously ignored or neglected. Interestingly, physical effects produced by such interactions have been experimentally observed \cite{ine}. Since our model involves certain constraints among the different terms of the Hamiltonian, the question of to what extent it describes a realistic BEC remains open. This is the question that we address here.

In this paper we show that the exactly solvable model of the two-mode BEC introduced in \cite{us,us2} can be derived from the microscopical description of a double-well BEC.  Showing this simply involves a generalization of the modes commonly employed in the two-mode approximation. The microscopic calculation we carry out here yields directly the diagonalizable model plus extra-terms that can be treated perturbatively as long as the overlap between the quasi localised modes in each well is much smaller than the trap asymmetry. We find that this is indeed the case in a wide variety of experiments involving double-well BECs. The behaviour of the diagonalizable model under generic perturbations has already been studied in \cite{mann}. Exploiting these results we compute the relevant corrections of the eigenstates within perturbation theory. Thus, we provide a complete toolbox for the analytical description of a wide range of double-well BECs, which includes not only the ground state of the system but all the eigenvalues and eigenvectors.
\begin{figure}[t!]
\includegraphics[width=\linewidth]{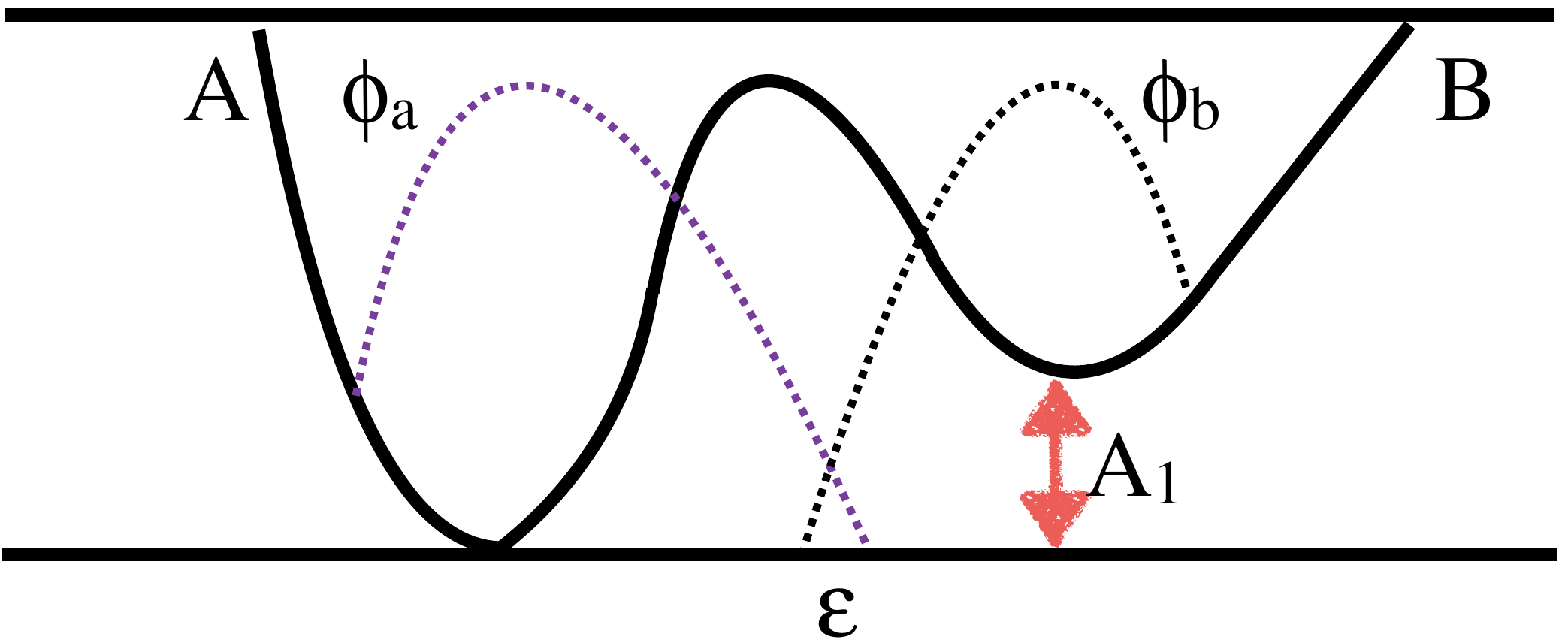}
\caption{An asymmetric double-well BEC consisting of wells A and B with energy offset $A_1$.  $\phi_a$, $\phi_B$ are almost-localised modes with a small overlap characterised by $\epsilon$ (see the text). In the Bose-Hubbard model, the two-mode approximation considers two equally balanced superpositions of $\phi_a$ and $\phi_b$. Here, we consider a more general superposition (see Eq. (\ref{eq:lmodes}) which allows to treat the Hamiltonian analytically, as long as $\epsilon<<A_1$.} \label{fig:sketch}
\end{figure}

%In the case of an asymmetric double-well Bose-Einstein condensate,  this extra term corresponds to a perturbation in the single-particle free energy and the tunneling term between wells.  Such perturbations have already been studied in \cite{mann} yielding very small corrections to the exact  analytical model.  We find that the perturbative term is not negligible when the well asymmetry is very small.  However, for larger asymmetries this term can be ignored.\\
Let us discuss our results in more detail. We start with the microscopic description of a BEC. Consider the many-body energy functional for bosonic particles of mass $m$ trapped in a potential $V(\mathbf{r})$ undergoing two-body collisions with s-wave scattering length $a$:
\begin{eqnarray} \label{eq:hamiltonian}
\hat{H}&=&\hat H_1+\hat H_2,\nonumber\\
\hat H_1&=&\int d\mathbf{r} \big(-\frac{\hbar^2}{2m}\hat\Psi^{\dagger}\nabla^2\hat\Psi+\hat\Psi^{\dagger}V(\mathbf{r})\hat\Psi\big),\nonumber\\
&=&\int d\mathbf{r} \hat\Psi^{\dagger}H_t\hat\Psi,\nonumber\\
H_{2}&=&\frac{g}{2}\int d\mathbf{r}\hat\Psi^{\dagger}\hat\Psi^{\dagger}\hat\Psi\hat\Psi,
\end{eqnarray}
where  $g=\frac{4\pi\hbar^2a}{m}$ is the coupling strength and $H_t$ is the  Hamiltonian of the trap.  The wavefunction $\hat\Psi$ can be expanded in terms of the eigenfunctions $\phi_i$ of the total Hamiltonian $\hat H$  and their corresponding annhilation operators $\hat c_i$  as 
\begin{equation}\label{eq:modes}
\hat\Psi=\sum_i\phi_i\hat c_i. 
\end{equation}
At absolute zero temperature and in the absence of particle collisions, all particles occupy the ground level of the single-particle Hamiltonian $\hat H_1$ producing a BEC.  In this case, the ground-state wavefunction $\phi_S$ is symmetric.  However, in the case  that particles collide with interaction energy small compared to the single-particle energy,  we can consider that particles occupy the two lowest eigenstates of the  Hamiltonian, i.e the antisymmetric mode $\phi_A$ is also populated. In this case, a two-mode approximation is valid and the total wavefunction of the system is well described by  
\begin{equation}\label{eq:2modes}
\hat\Psi=\phi_1\hat c_1+\phi_2\hat c_2. 
\end{equation}
The real wavefunctions $\phi_1$ and $\phi_2$  are orthonormal modes $\int dr \phi_{i}\phi_{j}=\delta_{i,j}$ with $i,j=1,2$ which have a small transition amplitude such that  
\begin{equation}\label{eq:overlap1}
E_{12}=\int dr\phi_1H_t\phi_2. 
\end{equation}
Typically, the modes in terms of the symmetric and antisymmetric solutions are written as,
\begin{eqnarray}\label{eq:almostmodes}
\phi_1&=&\frac{1}{\sqrt{2}}(\phi_S+\phi_A),\nonumber\\
\phi_2&=&\frac{1}{\sqrt{2}}(\phi_S-\phi_A).\nonumber\\
\end{eqnarray}
By substituting the two-mode approximation Eq. (\ref{eq:almostmodes}) in the Hamiltonian (\ref{eq:hamiltonian}) we obtain,
\begin{eqnarray}\label{hamiltonian}
H_1&=&E_1c_1^\dagger c_1+E_2c_2^\dagger c_2+ E_{12}(c_1^{\dagger}c_2+c_2^{\dagger}c_1)\nonumber\\
H_2&=&U_{1111}c_1^{\dagger 2} c_1^{2}+U_{2222}c_2^{\dagger 2} c_2^2+ +4U_{1212}c_1^{\dagger }c_2^{\dagger }c_1c_2\nonumber\\&+&2U_{1112}(c_1^{\dagger 2}c_1c_2+h.c)+2U_{2212}(c_2^{\dagger 2}c_1c_2+h.c)\nonumber\\&+& U_{1122}(c_1^{\dagger 2}c_2^2+h.c)\nonumber\end{eqnarray}
where 
\begin{eqnarray}\label{eq:terms}
E_1&=&\int dr\phi_1H_t\phi_1\,,\,E_2=\int dr \phi_2H_t\phi_2,\nonumber\\ 
U_{1111}&=&\int dr \phi^{ 2}_1\phi_1^2\, ,\, U_{2222}=\int dr\phi^{ 2}_2\phi_2^2\nonumber\\  
U_{1212}&=&\int dr|\phi_1|^{ 2}|\phi_2|^2\, ,\,  U_{1112}=\int dr\phi^{ 2}_1\phi_1\phi_2 \nonumber\\  U_{1222}&=&\int dr\phi^{2}_2\phi_1\phi_2 \, ,\, U_{1122}=\int dr\phi^{ 2}_1\phi_2^2.
\end{eqnarray}   

The Hamiltonian Eq. (\ref{hamiltonian})describes a constant number of particles $N=n_1+n_2$ where 
\begin{equation}\label{eq:number}
n_{1,2}=c_{1,2}^{\dagger}c_{1,2} 
\end{equation}
of them occupy mode $_{1,2}$  with corresponding energy $E_{1,2}$.  The number of particles in each mode change with probability amplitude $E_{12}$.  In addition to this, particles undergo  same-mode collisions with scattering lengths $U_{1111}$ and $U_{2222}$, elastic collisions between particles in different modes $U_{1122}$  and mode-exchange collisions $U_{1112}$, $U_{1222}$, $U_{1122}$. While the latter terms are usually neglected, their effects have been observed in the laboratory \cite{ine} and they have been extensively discussed theoretically \cite{us2}. 

Let us now focus in double-well BECs and distinguish two cases of interest, namely symmetric and asymmetric potentials. We label the symmetric and antisymmetric solutions  of the single particle Hamiltonian corresponding to the symmetric well as $\phi_A^s, \phi_S^s$  and
$\phi_A^a, \phi_S^a$ the solutions in the asymmetric potential. Therefore, in the case of the symmetric potential the two-mode approximation corresponds to two almost-localized modes
\begin{eqnarray}\label{eq:symmetricmodes}
 \phi_1^s=\frac{1}{\sqrt{2}}(\phi_S^a+\phi_A^a)\nonumber\\
 \phi_2^s=\frac{1}{\sqrt{2}}(\phi_S^a-\phi_A^a).
 \end{eqnarray}
  In the literature of double-well BECs \cite{cirac}, these modes are commonly labeled 
  \begin{equation}\label{eq:symmetricmodes2}
  \phi^s_1=\phi_a\,;\,  \phi^s_2=\phi_b 
  \end{equation}
  to emphasize that they are almost localized modes in well A and B respectively. We will use this notation here as well. 

In the asymmetric potential case, we can write the modes $\phi_1^a$ and $\phi_2^a$ in terms of the symmetric quasilocalized modes,
\begin{eqnarray}\label{eq:lmodes}
\phi_1^a&=&\cos({\theta}/{2})\phi_a-\sin({\theta}/{2})\phi_b\nonumber\\
\phi_2^a&=&\cos({\theta}/{2})\phi_b+\sin({\theta}/{2})\phi_a\nonumber
\end{eqnarray}
It is easy to see that in terms of the symmetric and antisymmetric solutions of the symmetric potential we obtain
\begin{eqnarray}\label{eq:almostmodes}
\phi_1^a&=&\cos\Omega\phi_A^s+\sin\Omega\phi_S^s\nonumber\\
\phi_2^a&=&-\sin\Omega\phi_A^s+\cos\Omega\phi_S^s\nonumber\\
\end{eqnarray}
where $\Omega=\frac{1}{2}(\theta+\frac{\pi}{2})$.  Note that we recover the symmetric case when $\theta=0$.

We have introduced the set of quasilocalized modes because the integrals in the Hamiltonian are easily evaluated in the $\phi_a$ and $\phi_b$ basis.  
%We can show that  $\int\phi^3_a\phi_b<<\int\phi^3_1\phi_2$ using that $\int(\phi_A^s)^3\phi_S^s=0$.  
They are nearly normalized modes with  $\int dr\phi_a\phi_a=1+\epsilon$, $\int dr\phi_b\phi_b=1-\epsilon$ , where the amplitude of transition between them 
\begin{equation}\label{eq:epsilon}
\epsilon=\int\,dr\,\phi_aH_t\phi_b 
\end{equation}
is assumed to be very small.  Therefore, all the fourth-order integrals containing both modes, such as :
\begin{eqnarray}\label{eq:ints}
 I_1=\int dr\phi_b^2\phi_a^2\,,\,I_2=\int dr\phi_a^3\phi_b,\,,\,I_3=\int dr\phi_b^3\phi_a 
\end{eqnarray}
are  of order $\mathcal{O}(\epsilon^2)$. 

Relaxing the notation, the modes of the asymmetric potential are related with the quasilocalized modes by the transformation:
\begin{eqnarray}\label{eq:lmodes}
\phi_1&=&\cos({\theta}/{2})\phi_a-\sin({\theta}/{2})\phi_b\nonumber\\
\phi_2&=&\cos({\theta}/{2})\phi_b+\sin({\theta}/{2})\phi_a.
\end{eqnarray}

Note that for the case $\theta=0$, the two-mode approximation becomes  
\begin{equation}\label{eq:theta0}
\hat\Psi=\phi_a\hat a+\phi_b\hat b,  
\end{equation}
which yields a Hamiltonian which describe two non-interacting well-localised modes,
\begin{eqnarray}\label{hamiltonian}
H&=&E_aa^\dagger a+E_bb^\dagger b
+U_{aa}a^{\dagger 2}a^{2}+U_{bb}b^{\dagger 2} b^2\nonumber\end{eqnarray}
with 
\begin{eqnarray}\label{eq:parameters}
E_{a}=\int dr\phi_aH_t\phi_a\, ,\,  U_{aa}=\int dr\phi_a^4, \nonumber\\
E_{b}=\int dr\phi_bH_t\phi_b\, , \, U _{bb}=\int dr\phi_b^4. 
\end{eqnarray}
In this case the modes $\phi_a$ and $\phi_b$ become  orthonormal. 
Another case of interest is that of $\theta=\pi/2$. In this case we obtain the two-mode approximation commonly used in the literature for a symmetric double well potential which leads to the well known  
Hubbard-Bose Hamiltonian after neglecting the terms corresponding to mode-exchange collisions $U_{1212}$,  $U_{1112}$, $U_{2212}$ and $U_{1122}$.  

Here we consider a more general mode decomposition parametrized by the angle $\theta$. However, this simple change has strong consequences. After evaluating the integrals in this new set of modes and adding the constant $-N$ we obtain the Hamiltonian,
\begin{equation}\label{diagonalizable}
H=\mathcal{H}_0+\mathcal{H}^{'}+\mathcal{H}^{''}
\end{equation}
where $\mathcal{H}_0$ is diagonalizable: \cite{us,us2}
\begin{eqnarray}\label{eq:lapartediag}
\mathcal{H}_0&=&A_1\cos\theta(c_1^\dagger c_1-c_2^\dagger c_2)+ A_1\sin\theta(c_1^{\dagger}c_2+c_2^{\dagger}c_1)\nonumber\\
&+&A_2(1+\cos^2\theta)(c_1^{\dagger 2} c_1^{2}+c_2^{\dagger 2} c_2^2) +4A_2\sin^2\theta c_1^{\dagger }c_2^{\dagger }c_1c_2\nonumber\\&+&2A_2\cos\theta\sin\theta(c_1^{\dagger 2}c_1c_2-c_2^{\dagger 2}c_1c_2+h.c)\nonumber\\&+& A_2\sin^2\theta(c_1^{\dagger 2}c_2^2+h.c)\end{eqnarray}
 and $\mathcal{H}^{'}$, $\mathcal{H}^{''}$ are of order $\mathcal{O}(\epsilon)$, $\mathcal{O}(\epsilon^2)$ respectively. 
Here 
\begin{equation}\label{trapasimmetry}
A_1=\frac{1}{2}(E_a-E_b) 
\end{equation}
and 
\begin{equation}\label{eq:a2}
A_2=\frac{1}{2}U 
\end{equation}
for $U=U_{aa}=U_{bb}$ assuming that the scattering length of particles in each mode is equal. Close attention must be taken in evaluating $A_1$ and $A_2$  since the wavefunctions must be renormalized before the integration is carried out.  $\mathcal{H}_0$ describes a BEC in an asymmetric double-well potential. $c_1^{\dagger}c_1$ and $c_2^{\dagger}c_2$ correspond to the number of particles in each well. The wave functions $\phi_1$ and $\phi_2$ overlap giving rise to tunneling of particles through the potential barrier. Particles undergo on-site collisions $c^{\dagger}_{1,2}c^{\dagger}_{1,2}c_{1,2}c_{1,2}$ and  in the overlapping region of the wavefuntions particles of different wells can collide $c^{\dagger}_{1}c^{\dagger}_{2}c_{2}c_{1}$. Two other interesting effects are present in the Hamiltonian. One of them in coherent tunneling in which two particles collide and tunnel as a single particle $c^{\dagger}_{1,2}c^{\dagger}_{1,2}c_{2,1}c_{2,1}$. These effects has already been observed in laboratory. A less known effect is collision assisted tunneling in which a particle tunnels thanks to energy gained during a on-site collision $c^{\dagger}_{1}c^{\dagger}_{1}c_{1}c_{2}$. 

The eigenstates of the Hamiltonian $\mathcal{H}_0$ are 
\begin{equation}\label{eq:eigenstates}
|\Phi_{n_1,n_2}\rangle=\exp(\frac{\theta}{2}(c_1^{\dagger}c_2-c_1c_2^{\dagger}))|n_1,n_2\rangle
\end{equation}
 with corresponding eigenenergy 
 \begin{equation}\label{eq:eigenenergy}
 E_{n_1,n_2}=A_1(n_1-n_2)+A_2(n_1-n_2)^2. 
 \end{equation}
 Note also that the two-mode approximation requires that on-site interactions are much stronger than the interactions between particles in different wells. This is possible if we restrict ourselves to small values of $\theta$ such that 
 \begin{equation}\label{eq:theta}
 \sin{\theta}\simeq \theta<<1\, , \,\cos{\theta}\simeq 1.
 \end{equation}

Let us consider the term $\mathcal{H}^{'}$, 
\begin{equation}\label{eq:firstorder}
\mathcal{H}^{'}=-\epsilon(\frac{\theta}{2}(c_1^{\dagger}c_1-c_2^{\dagger}c_2)+c_1^{\dagger}c_2+c_2^{\dagger}c_1),
\end{equation}
where we have already assumed the approximation in Eq. (\ref{eq:theta}). 

By comparing Eqs. (\ref{eq:lapartediag}) and (\ref{eq:firstorder}), we see that we can treat $\mathcal{H}^{'}$ as a perturbation, as long as $\epsilon << A_1\theta$. This is indeed the case in a wide variety of experiments involving double-well condensates \cite{spatially, doublewell1,doublewell2}. The tunnelling rate $E_{12}\simeq A_1\theta+\epsilon$ takes experimental values ranging from $5\cdot10^{-4}\,\operatorname{Hz}\times h$ \cite{doublewell1} to $2\,\operatorname{Hz}\times h$ \cite{spatially}, while the energy offset between the wells can be as high $A_1=530\,\operatorname{Hz}\times h$ \cite{doublewell2}. Even if the wells are intended to be perfectly symmetric, the uncertainty in the trap depth leads us to assume a minimum trap asymmetry of $A_1\simeq20\,\operatorname{Hz}\times\,h$ \cite{spatially}. 

Using the results in \cite{mann}, we find the following non-vanishing matrix elements in perturbation theory:
\begin{eqnarray}\label{eq:nonvanishing}
\langle \Phi_{n_1\,n_2}|H^{'}|\Phi_{n_1\,n_2}\rangle&=&-\frac{3}{2}\epsilon\,\theta\,(n_1-n_2)\nonumber\\
\langle \Phi_{n_1+1\,n_2-1}|H^{'}|\Phi_{n_1\,n_2}\rangle&=&-\epsilon\sqrt{n_2\,(n_1+1)}\nonumber\\
\langle \Phi_{n_1-1\,n_2+1}|H^{'}|\Phi_{n_1\,n_2}\rangle&=&-\epsilon\sqrt{n_1\,(n_2+1)}.
\end{eqnarray}
%
%Therefore, in the case that   $\epsilon=\int\phi_aH_t\phi_b$ can be neglected we obtain from first principals,  a Hamiltonian which has exact anaytical solution.  In the case $\epsilon$ is small but not negligelble, we have a perturbation to the diagonalizable Hamiltonian which has already been studied in \cite{man}.  The Hamiltonian (\ref{diagonalizable}) and it's applications to  Bose-Einstein condensates has been studied in detial in \cite{fuentes-barberis} and \cite{barberis-fuentes}.\\
%Note that the discussion is independent of the trap's Hamlitonian $H_t$.  We only requiered that the Hamitonian is such that a two-mode approximation is valid and that $\epsilon$ is small.  Here we have focused our discussion to a particular problem of this type: the asymmetric double-well potential. \\

Now let us discuss the term $\mathcal{H}^{''}$, which includes all the fourth-order integrals containing both modes:
\begin{eqnarray}\label{eq:secondorder}
\mathcal{H}^{''}&=&4\,I_1 c_1^{\dagger }c_2^{\dagger }c_1c_2\,+2\,I_2(c_1^{\dagger 2}c_1c_2+h.c)\nonumber\\
&+&2\,I_3(c_2^{\dagger 2}c_1c_2+h.c)+I_1(c_1^{\dagger 2}c_2^2+h.c)+\mathcal{O}(\theta),
\end{eqnarray}
where the $\mathcal{O}(\epsilon^2)$ integrals $I_1$, $I_2$, $I_3$ are defined in Eq. \ref{eq:ints}. The leading contributions to the non-vanishing matrix elements of the perturbation are:
\begin{eqnarray}
\langle \Phi_{n_1\,n_2}|H^{'}|\Phi_{n_1\,n_2}\rangle&=&4\,n_1\,n_2\,I_1\nonumber\\
\langle \Phi_{n_1+1\,n_2-1}|H^{'}|\Phi_{n_1\,n_2}\rangle&=&2\,(I_2+\,I_3) \sqrt{n_2\,(n_1+1)}\nonumber\\
\langle \Phi_{n_1-1\,n_2+1}|H^{'}|\Phi_{n_1\,n_2}\rangle&=&2\,(I_2+\,I_3) \sqrt{n_1\,(n_2+1)}\nonumber\\
\langle \Phi_{n_1+2\,n_2-2}|H^{'}|\Phi_{n_1\,n_2}\rangle&=&I_1\sqrt{(n_1+1)\,(n_1+2)\,n_2\,(n_2-1)}\nonumber\\
\langle \Phi_{n_1-2\,n_2+2}|H^{'}|\Phi_{n_1\,n_2}\rangle&=&I_1\sqrt{(n_2+1)\,(n_2+2)\,n_1\,(n_1-1)}\nonumber
\end{eqnarray}
The behaviour of several physical quantities of interest in a model described by a Hamiltonian $\mathcal{H}_0$ under generic perturbations with the same mathematical structure as $\mathcal{H}^{
'}$ and $\mathcal{H}^{''}$ was thoroughly discussed in \cite{mann}.

%is negligeble when the trap asymmetry $A_1$ is much larger than the overlap  $\int\phi_aH_t\phi_b$.  This situations cannot be satisfied when the well's asymmetry is very small. 

In summary, we introduce a toolbox for the analytical description of a two-mode BEC. 
Our results are found by means of a generalisation of the modes commonly used in the two-mode approximation.  The Bose-Hubbard model corresponds to a particular case of the modes we introduce here. An important advantage of our scheme is that it includes mode-exchange and coherent tunnelling interaction terms that have been observed experimentally and are typically neglected.  Moreover, while the Bose-Hubbard hamiltonian is commonly solved numerically, our model can be fully treated with analytical methods. We not only find the ground state but the full spectrum and the eigenstates. We believe that this will be of great benefit for quantum information and metrology application with BEC setups. As a first example of interest, our discussion has been focused on the asymmetric double-well BEC. However, the theory we have here developed can be applied to a more general situation since the only assumptions we have made are that the condensate wavefunction can be approximated by two modes and that the transition amplitude between them is very small. For instance,  a case of interest is a single-well Bose-Einstein condensate consisting of atoms in two hyperfine levels \cite{JILA2}. In such case, mode-exchange collisions  have been predicted by microscopic calculations \cite{julienne}. The exactly solvable model can be generalized to the multi-mode case (including optical lattices) and many-body interactions. It is therefore of great interest to explore as well the microscopic derivation of those cases.  
\section*{Acknowledgements}
IF and CS acknowledge funding from EPSRC (CAF Grant No. EP/G00496X/2 to I. F.)

\end{document}